\documentclass[aps, amsfonts,amsmath,showpacs,nofootinbib]{revtex4}

 \ifx\pdfoutput\undefined
   \usepackage[dvips]{graphicx}
   \else
   \usepackage[pdftex]{graphicx}
   \fi
  \usepackage{epstopdf}

\def\be{\begin{equation}}
\def\ee{\end{equation}}
\def\beq{\begin{equation}}
\def\eeq{\end{equation}}
\def\bea{\begin{eqnarray}}
\def\eea{\end{eqnarray}}
\def\bml{\begin{subequations}}
\def\blea{\bml\begin{eqnarray}}
\def\elea{\end{eqnarray}\end{subequations}}

\begin{document}

\title{Towards a kinetic theory of strings}

\author{Vitaly Vanchurin}

\email{vanchurin@stanford.edu}

\affiliation{Department of Physics, Stanford University, Stanford, CA 94305}

\begin{abstract}

We study the dynamics of strings by means of a distribution function $f({\bf A}, {\bf B}, {\bf x}, t)$ defined on a 9+1D phase space, where ${\bf A}$ and ${\bf B}$ are the correlation vectors of right- and left-moving waves. We derive a  transport equation (an analogous to Boltzmann transport equation for particles) that governs the evolution of long strings with Nambu-Goto dynamics as well as reconnections taken into account. We also derive a system of coupled transport equations (an analogous to BBGKY hierarchy for particles) which can simultaneously describe long strings $\tilde{f}({\bf A}, {\bf B}, {\bf x}, t)$ as well as simple loops $\mathring{f}({\bf A}, {\bf B}, {\bf x}, t)$ made out of four correlation vectors. The formalism can be used to study non-linear dynamics of fundamental strings, D-brane strings or field theory strings. For example, the complicated semi-scaling behavior of cosmic strings translates into a simple solution of the transport system at small energy densities.

\end{abstract}


\maketitle

\section{Introduction}

The semiclassical evolution of long fundamental strings $l \gg R \gg l_s$ (where $R$ is the size of compact dimensions and $l_s$ is string length) is described fairly accurately by Nambu-Goto action until the strings start to intersect in the target space. Such intersection can lead to reconnections with non vanishing probability $p > 0$ which significantly complicates the dynamics.  The problem proved to be relevant not only for  fundamental strings or D-brane strings, but also for field theory strings on various gravitational background. In fact most of numerical simulations of strings were designed to study cosmic strings with $p \approx 1$ \cite{BB1,AS1,BB2,VHS, VOV, MS,RSB}. The numerical studies revealed a very interesting semi-scaling phenomena: scaling of large scales and non-scaling of small scales \cite{VOV2, OV, BPOS} which one can understand reasonably well even analytically \cite{PR, DPR, Vanchurin1, Vanchurin2, Vanchurin3}. 

In addition to numerical simulations a couple of attempts where made to describe the dynamics of strings by means of a transport equation on a probability distribution of only string lengths  \cite{LT, LT2} above Hagerdon temperature or on a probability distributions of string intersections with a given two dimensional surface \cite{BI}. Unfortunately, not all of the transport phenomena can be described in terms of these probability distributions. In this paper we describe the statistics of strings with a probability distribution of the left- and right-moving correlation vectors (representing diamond-shaped section of the world-sheet) which are directly related to observable quantities. The phase space remains reasonably small (only nine dimensional) but all of the relevant transport phenomena (such as Nambu-Goto evolution, transverse and longitudinal reconnections, production of loops and background gravitational effects) can be included in the dynamics.

The paper is organized as follows. In the next section we define correlation vectors and construct the corresponding phase space. In Sec. \ref{transport} the transport equation is derived for long strings with Nambu-Goto evolution and transverse collisions taken into account. The dynamics of a system with long strings and simple loops is analyzed in Sec. \ref{hierarchy}. The main results are summarized in the conclusion. 

\section{Correlation vectors} \label{correlations}

The classical dynamics of strings in the limit of zero coupling $g_s=0$ is described by Nambu-Goto action:
\be
S_{NG} = - \frac{1}{2\pi \alpha'} \int \sqrt{-{\det}(h_{ab})} d\sigma dt  
\ee
where $h_{ab} = g_{\mu\nu} x^\mu_{,a}x^\mu_{,b}$. For convenience we also assume that  $2 \pi \alpha'= 1$ and $l_s=\frac{1}{\sqrt{2 \pi}}$. The corresponding equation of motion is
\be
x_{,a}^{\mu\, ;a} + \Gamma^{\mu}_{\nu \tau} h^{ab} x^\nu_{,a}x^\tau_{,b} =0.
\label{eq:EOM}
\ee
After gauge fixing ($h_{01} =0$, $h_{00}+h_{11}=0$, and $t=x^0$) on a flat background we get a system of equation for three-vectors
\be
\ddot{\bf x} - {\bf x}'' = 0, \;\;\;\;\;\;\;\;\;\;\;\;\;\;\;\;\;\;\;\;\;\;\;\;\dot{\bf x}^2 + {\bf x}'^2 = 0, \;\;\;\;\;\;\;\;\;\;\;\;\;\;\;\;\;\;\;\;\;\;\;\;{\bf x}' \cdot \dot{\bf x} = 0,
\ee
whose solution can be described in terms of left-  and right-moving waves
\be
\bold{x}(\sigma, t) = \frac{\bold{a}(\sigma- t) +\bold{b}(\sigma + t) }{2}
\label{eq:wave}
\ee
with condition $|\bold{a}' |=|\bold{b}' |=1$. 

In general the exact evolution of a given infinitesimal segment at $\sigma_0$ is known only if all of the functions $\bold{a}(\sigma)$ and $\bold{b}(\sigma)$ for all strings are known. In a kinetic theory such evolution should be describable by a system of equations analogous to the BBGKY hierarchy for interacting particles. Truncating the system at a giving order is equivalent to neglecting higher order correlations which is often a starting point in the kinetic theory. At the leading order the truncation of BBGKY hierarchy for particles leads to the molecular chaos assumption such that interacting particles are uncorrelated. Following the same logic for strings we define an ensemble of possible continuations of ${\bf a}(\sigma)$ and ${\bf b}(\sigma)$ away from $\sigma_0$ described by correlation vectors
\bea
{\bf A}(\sigma_0) \equiv - \left \langle  \int_{-L/2}^{L/2}  d\sigma {\bf a}'(\sigma_0+\sigma) \right \rangle
\;\;\;\;\;\;\;\;\;\;\;\;\;\text{and}\;\;\;\;\;\;\;\;\;\;\;\;\;\;
{\bf B}(\sigma_0) \equiv \left \langle \int_{-L/2}^{L/2}  d\sigma {\bf b}'(\sigma_0+\sigma) \right \rangle
\label{eq:define_A} 
\eea
for sufficiently large $L \gg A \equiv |{\bf A}|$, $L \gg B \equiv |{\bf B}|$ (Note a minus sign in the definition of ${\bf A}$.)  The exact value of $L$ is unimportant and one can safely send $L \rightarrow \infty$ for the correlation vectors defined on infinite strings. Describing the ensemble of continuations with correlation vectors and ignoring higher statistical momenta is an assumption analogous to the molecular chaos assumption for particles. We will refer to it is as the strings chaos assumption. 
 
The string chaos assumption is equivalent to approximating left- and right-moving waves by piece-wise linear functions such that neighboring segments are completely uncorrelated. Then the world-sheet can be viewed as a collection of diamond shaped regions glued together to form a curved surface and an arbitrary point with correlation vectors $\bf A$ and $\bf B$ would be located somewhere on the diamond $(\bf A, B)$. In the limit where the string coupling is small  (but non-zero) the Nambu-Goto action gives a good description of the dynamics of long strings $l \gg R \gg  l_s$ up until the point when strings start to pass through each other in the target space. A possible outcome of such intersection includes reconnections of right- and left-moving waves with non-zero probability  whose exact value would depend on compacification details. The probability is typically small for fundamental or D-brane strings $p \ll 1$ but can be quite large for field theory strings $p\approx 1$. In either case the system with a large number of strings becomes quickly intractable, but one might hope to understand the dynamics by means of a kinetic theory. If one applies the strings chaos assumption to the entire network and approximates all of the strings by piecewise linear functions then the entire world-sheet of all strings including all reconnections would be approximated by a collection of diamonds. Then the main problem is to describe the statistical evolution of the system with a transport equation (analogous to the Botzmann transport equation for particles). 

\section{Transport equation}\label{transport}

Let us define the energy density (or invariant length density) of strings  $f({\bf A}, {\bf B}, {\bf x}, t)$ with correlation vectors ${\bf A}$ and ${\bf B}$ at spatial position ${\bf x}$ and time $t$.  Then the evolution of this distribution from time $t$ till time $t+\delta t$ is given by
\be
f({\bf A}, {\bf B}, {\bf x} +{\bf v}({\bf A},{\bf B}) \delta t, t + \delta t) = f({\bf A}, {\bf B}, {\bf x}, t) +  \left ( \frac{\partial f}{\partial t} \right )_{G}  \delta t + \left ( \frac{\partial f}{\partial t} \right )_{NG} \delta t + \left ( \frac{\partial f}{\partial t} \right )_{T} \delta t+ \left ( \frac{\partial f}{\partial t} \right )_{L} \delta t
\ee
where  the expected velocity of an infinitesimal segment due to the string chaos assumption is ${\bf v}({\bf A},{\bf B}) \equiv \frac{1}{2}\left ( \hat{\bf A} +\hat{\bf B} \right )$ and $\hat{\bf A} \equiv \frac{\bf A}{A}$,  $\hat{\bf B} \equiv \frac{\bf B}{B}$ are the unit vectors. The corresponding transport equation is
\be
\frac{\partial f}{\partial t} +  {\bf v} \cdot \frac{\partial f}{\partial {\bf x}} +  \hat{\bf H} f  = \left ( \frac{\partial f}{\partial t} \right )_{NG} + \left ( \frac{\partial f}{\partial t} \right )_{T}+ \left ( \frac{\partial f}{\partial t} \right )_{L} 
\label{eq:transport}
\ee
where the gravitational term is expected to be given by some differential operator
\be
\left ( \frac{\partial f}{\partial t} \right )_{G }  \equiv - \hat{\bf H} f
\ee
which depends on the background geometry  (see Appendix \ref{Friedmann}). The dependence of $f$, ${\bf v}$ and $\hat{\bf H}$ on ${\bf A}$, ${\bf B}$ and ${\bf x}$ will be suppressed throughout the paper for brevity of notations. The three "collision" terms on the right hand side of (\ref{eq:transport}) represent: Nambu-Goto evolution, transverse (i.e. intersection of nearby strings) and longitudinal (i.e. production of simple loops) reconnections. The exact role played by these terms will become more transparent once we derive them below.

To simultaneously describe the populations of long strings as well as small loops it will be convenient to define two distribution functions  $\tilde{f}$ and $\mathring{f}$ such that  $f = \tilde{f} + \mathring{f}$, however at very large energy densities it is expected that most of the energy is in long strings $f \sim \tilde{f}$. In this limit we can neglect the production of small loops $ \left ( \frac{\partial f}{\partial t} \right )_{L} \approx 0$ in (\ref{eq:transport}):
\be
\frac{\partial \tilde{f}}{\partial t} +  {\bf v} \cdot \frac{\partial \tilde{f}}{\partial {\bf x}} +  \hat{\bf H} \tilde{f}  = \left ( \frac{\partial \tilde{f}}{\partial t} \right )_{NG} + \left ( \frac{\partial \tilde{f}}{\partial t} \right )_{T},
\label{eq:transport_long}
\ee
and study the remaining terms (Nambu-Goto and transverse) one by one.

As the time passes the neighboring world-sheet diamonds exchange correlation vectors due to classical dynamics described by the Nambu-Goto action.   \begin{figure}[htbp]
   \begin{center}
   \includegraphics[width=4.5in]{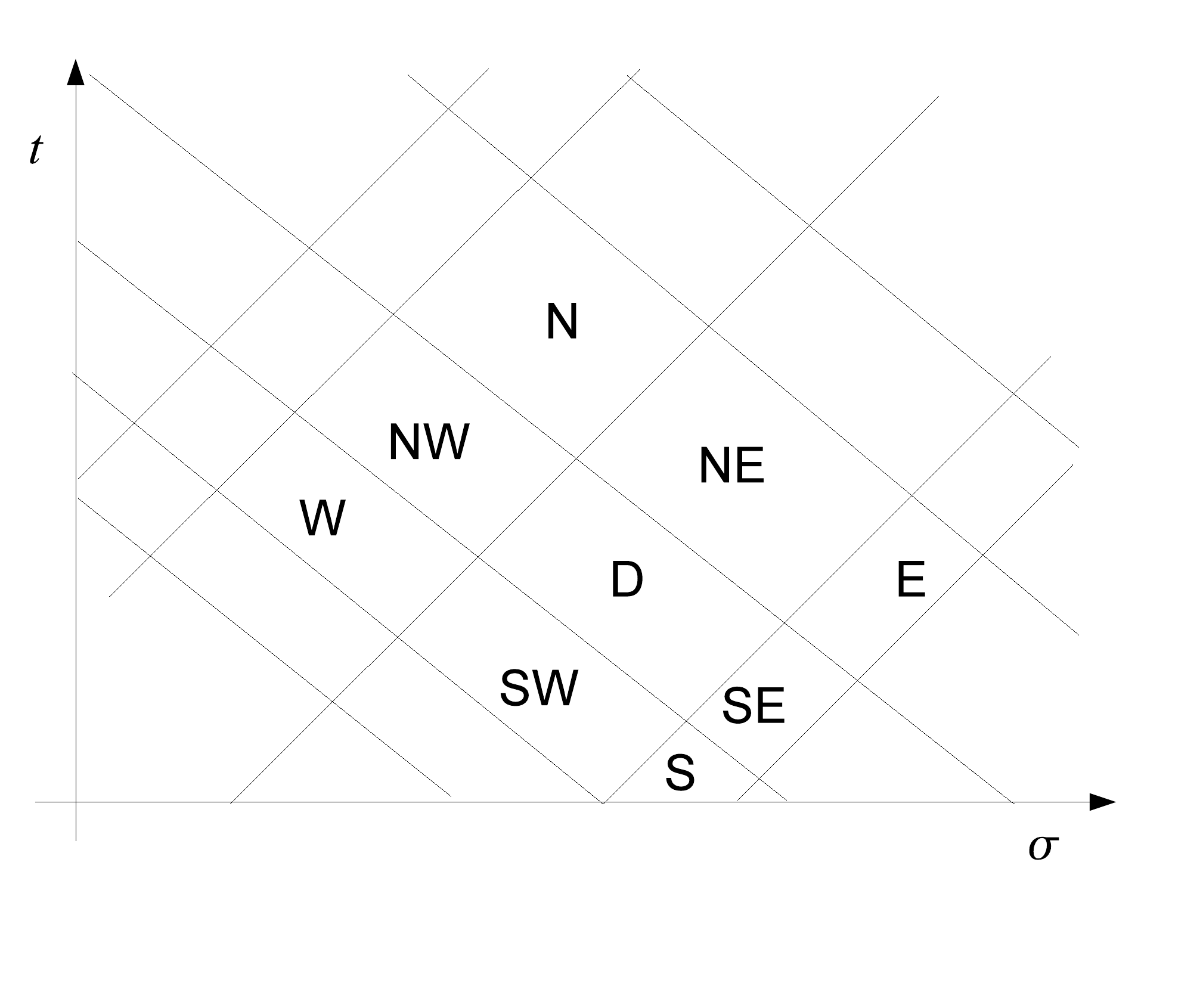}
   \caption{Diamond $D$ with all of its nearest neighbors $W, E, N, S, NW, NE, SW$ and $SE$.}
   \label{fig:neighbors}
   \end{center}
   \end{figure}Since the world-sheet area of a given diamond $D$ (see Fig.\ \ref{fig:neighbors}) with correlation vectors $\bf A$ and $\bf B$ is  $A B/2$, the probability of a random infinitesimal segment on $D$ to exit $D$ through $NW$ boundary in unit time is $\frac{A}{\sqrt{2}} \frac{1}{\sqrt{2}} \frac{2}{A B} = 1/A$ and the probability to exit $D$ through $NE$ boundary is $1/B$. Similarly a given infinitesimal segment on ether $SW$ or $SN$ diamonds can enter $D$ through $SW$ and $SN$ boundaries respectively. In the transport equation such evolution would be describe by two destruction and two creation terms representing transitions of infinitesimal segments through $NW, NE,  SE$ and $SW$ boundaries respectively, i.e.
\be
 \left ( \frac{\partial \tilde{f}}{\partial t} \right )_{NG}  =  - \frac{\tilde{f}({\bf A}, {\bf B})}{A}  -  \frac{\tilde{f}({\bf A}, {\bf B})}{B}  + \int d {\bf A}'  d {\bf B}' \frac{\tilde{f}({\bf A}', {\bf B}) \tilde{f}({\bf A}, {\bf B}') }{A A' \tilde{N}_A}  + \int d {\bf A}'  d {\bf B}' \frac{ \tilde{f}({\bf A}', {\bf B}) \tilde{f}({\bf A}, {\bf B}') }{B B' \tilde{N}_B} 
 \label{eq:nambu_long}
\ee
where 
\bea
\tilde{N}_A \equiv \int  \frac{d {\bf A} d {\bf B}}{A}  \tilde{f}({\bf A}, {\bf B})\;\;\;\;\;\;\;\;\;\;\;\;\;\;\;\;\;\;\;\;\;\;\;\;\tilde{N}_B \equiv \int  \frac{d {\bf A} d {\bf B}}{B}  \tilde{f}({\bf A}, {\bf B}).
\label{eq:normalization}
\eea

\begin{figure}[htbp]
   \begin{center}
   \includegraphics[width=4.5in]{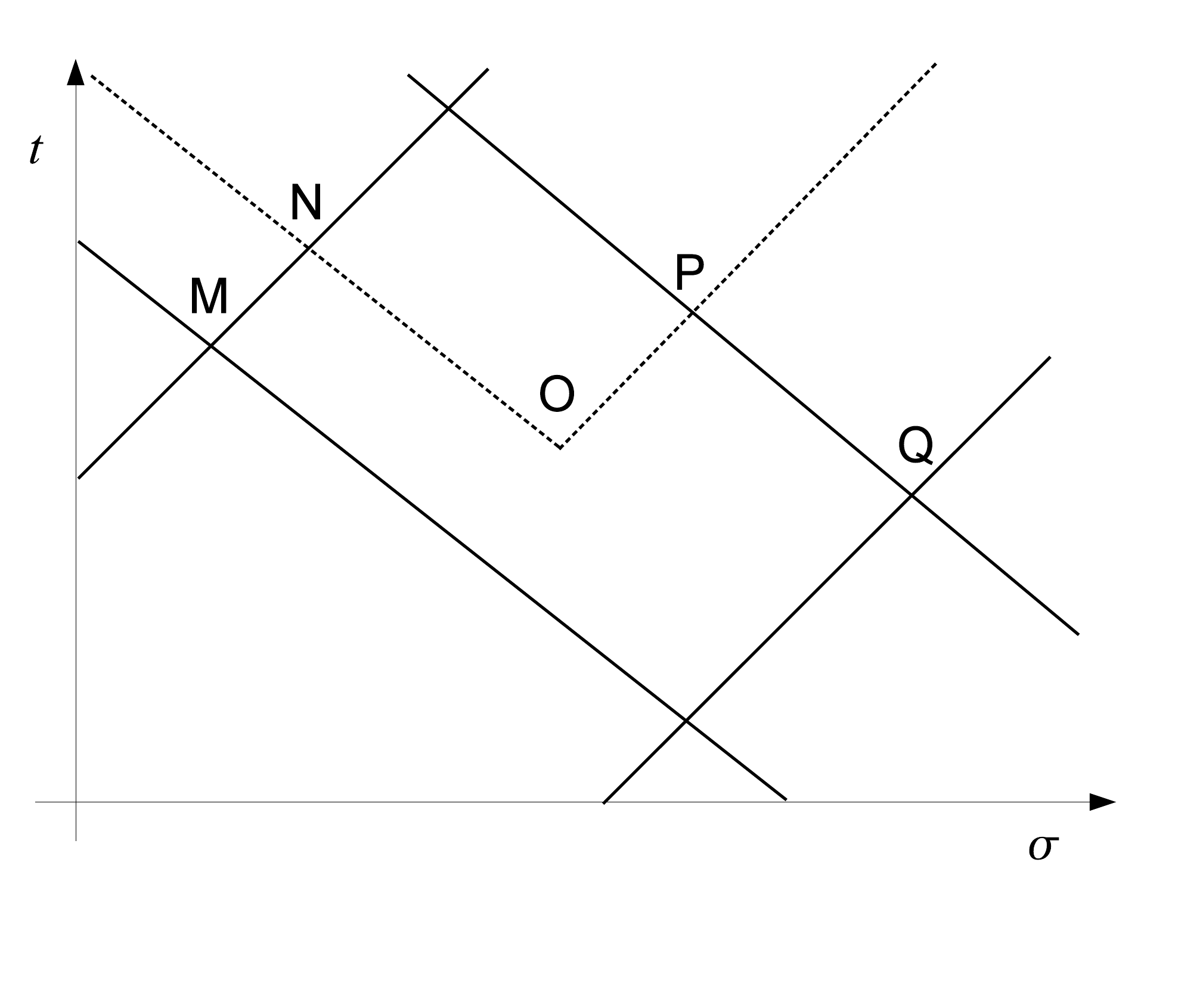}
   \caption{Diamond undergoing a transverse collisions at point $O$.}
   \label{fig:transverse}
   \end{center}
   \end{figure}

The transverse collisions describe interactions of diamonds which could be arbitrary far away on the world-sheet but nevertheless intersect in the target space. Such intersection might produce sudden deformations of the world-sheet with non-zero probability $p > 0$.  See Fig.\ \ref{fig:transverse}.   
%The area of a diamond's projection on a constant time surface is $|{\bf A} \times {\bf B}|/4$, and thus the probability for an infinitesimal segment to cross $MN$ or $OP$ per unit time is $\propto {|{\bf A} \times {\bf B}|/(4A)}$ and the probability to cross $PQ$ or $NO$ is $\propto |{\bf A} \times {\bf B}|/(4B)$. 
The area spanned by a diamond $({\bf A},{\bf B})$ is $\frac{AB}{4} |\hat{A} \wedge \hat{B}|$, where $\hat{A} \equiv (1, \frac{\bf A}{A})$,  $\hat{B} \equiv (1, \frac{\bf B}{B})$ are the null four-vectors. Therefore the probability for an infinitesimal segment to cross $MN$ or $OP$ ($PQ$ or $NO$) per unit time is $\propto 1/A$ ($\propto 1/B$) and the probability to collide with an infinitesimal segment of another diamond $({\bf A}',{\bf B'})$ is proportional to a four-volume given by $|(\hat{A}\wedge\hat{B})\wedge(\hat{A}'\wedge\hat{B}')|$. All of these transitions (across $MN$, $NO$, $OP$ and $PQ$ boundaries) lead to modification of the correlation vectors to be described by a collision term $ \left ( \frac{\partial \tilde{f}}{\partial t} \right )_{T} $ in the transport equation. Under the strings chaos assumption the colliding diamonds are taken to be completely uncorrelated suggesting the following transverse term 
\bea
 \left ( \frac{\partial \tilde{f}}{\partial t} \right )_{T}  = - p \int d {\bf A'}  d {\bf B'} \frac{\left ({A} + {B} \right )}{4} |(\hat{A}\wedge\hat{B})\wedge(\hat{A}'\wedge\hat{B}') | \tilde{f}({\bf A}, {\bf B})  \tilde{f}({\bf A'}, {\bf B'}) +  \label{eq:transverse} \;\;\;\;\;\; \;\;\;\;\;\; \;\;\;\;\;\; \;\;\;\;\;\; \;\;\;\;\;\; \;\;\;\;\;\; \;\;\;\;\;\; \;\;\;\;\;\; \;\;\;\;\;\; \;\;\;\;\;\; \;\;\;\;\;\; \;\;\;\;\;\; \;\;\;\; \\
p \int_{0 < \alpha,\beta < 1} d\alpha d {\bf A'}   d\beta d {\bf B'}  \frac{\left ({A} + {B} \right )}{4} \left ( |(\hat{A}\wedge \hat{B}')\wedge (\hat{A}'\wedge \hat{B})|  \tilde{f}({\bf A'}, \frac{\bf B}{\beta}) \tilde{f}(\frac{\bf A}{\alpha}, {\bf B'})  +  |(\hat{A}\wedge\hat{B})\wedge (\hat{A}'\wedge\hat{B}') |  \tilde{f}(\frac{\bf A}{\alpha}, \frac{\bf B}{\beta})  \tilde{f}({\bf A'}, {\bf B'}) \right) \;\;\; \notag
\eea
where the integration over $\alpha$ and $\beta$ describes an unknown location of the intersection on participating diamonds . The first and second lines in (\ref{eq:transverse}) represent respectively  the destruction and creation of infinitesimal segments due to transverse collisions. 

It follows that the transport equation (\ref{eq:transport}) together with Nambu-Goto (\ref{eq:nambu_long}) and transverse (\ref{eq:transverse}) terms takes the following simple form:
\bea
\frac{\partial \tilde{f}}{\partial t} +  {\bf v} \cdot \frac{\partial \tilde{f}}{\partial {\bf x}} + \hat{\bf H} \tilde{f} + \frac{\tilde{f}}{A} +  \frac{\tilde{f}}{B}   = \int d\alpha  d {\bf A'} d\beta  d {\bf B'} \left [ \left ( \frac{1}{A A' \tilde{N}_A} +  \frac{1}{B B' \tilde{N}_B}  \right ) \tilde{f}({\bf A'}, {\bf B}) \tilde{f}({\bf A}, {\bf B'})+ \;\;\;\;\;\;\;\;\;\;\;\;\;\;\;\;\;\;\;\;\;\;\;\;\;\;\;\;\;\;\;\;\;\;\;\;\;\;\;\;\;\;\;\;\;\; \right. \\ 
\left. +   p  \frac{\left ({A} + {B} \right )}{4} \left |\hat{A}\wedge\hat{B} \wedge \hat{A}'\wedge\hat{B}' \right | \left(\tilde{f}({\bf A'}, \frac{\bf B}{\beta}) \tilde{f}(\frac{\bf A}{\alpha}, {\bf B'}) +  \left (\tilde{f}(\frac{\bf A}{\alpha}, \frac{\bf B}{\beta}) -  \tilde{f}({\bf A}, {\bf B})  \right )  \tilde{f}({\bf A'}, {\bf B'}) \right )  \right ]. \notag
\label{eq:transport_gravity}
\eea
This equation can already be used to study networks of long strings  at large energy densities in Minkowski space (where $\hat{\bf H}=0$), but it is also possible to derive $\hat{\bf H}$ for more complicated background geometries. In the Appendix \ref{Friedmann} we provide a sample derivation of the operator $\hat{\bf H}$ for Friedmann backgrounds which is most relevant for cosmology. 

\section{Hierarchy of equations}\label{hierarchy}

Until now we have only concentrated on long strings, but it is straightforward to generalize our discussion to include simple loops. One can show that the simplest loops must contain at least two right-moving and two left-moving correlation vectors: ${\bf A}_1$, ${\bf A}_2$, ${\bf B}_1$ and ${\bf B}_2$ where ${\bf A}_1+{\bf A}_2  = {\bf B}_1+{\bf B}_2$  and  $A_1+A_2= B_1+B_2$ is the total invariant length of the loop. Such loops are describe by four diamonds $({\bf A}_1, {\bf B}_1)$,  $({\bf A}_1, {\bf B}_2)$,  $({\bf A}_2, {\bf B}_1)$ and  $({\bf A}_2, {\bf B}_2)$ that would overlap in the target space if the left- and right-moving waves were peciewise linear made out of $\{{\bf B}_1, {\bf B}_2\}$  and $\{{\bf A}_1, {\bf A}_2\}$ respectively. One might argue that this corresponds to an infinite number of self-intersections, but as we will see the longitudinal collisions of long strings produce non-self-intersecting loops which are only approximated with four correlation vectors under the string chaos assumption.

 More generally one might also want to include populations of more complicated loops consisting of 5, 6 and more correlation vectors with additional distribution functions and additional transport equations. A complete system of such equations would be analogous to the BBGKY hierarchy for particles, but we shall truncate the hierarchy at only two equations describing long strings  $\tilde{f}$ and the simplest loops $\mathring{f}$ made out of 4 correlation vectors:
 \bea
\begin{cases}
\frac{\partial \tilde{f} }{\partial t} +  {\bf v} \cdot \frac{\partial \tilde{f} }{\partial {\bf x}} + \hat{\bf H} \tilde{f} =  \left ( \frac{\partial \tilde{f}}{\partial t} \right )_{NG} +  \left ( \frac{\partial \tilde{f}}{\partial t} \right )_{T} +  \left ( \frac{\partial \tilde{f}}{\partial t} \right )_{L}  \\
\frac{\partial \mathring{f}}{\partial t} +  {\bf v} \cdot \frac{\partial \mathring{f}}{\partial {\bf x}} + \hat{\bf H} \mathring{f} =  \left ( \frac{\partial \mathring{f}}{\partial t} \right )_{NG} +  \left ( \frac{\partial \mathring{f}}{\partial t} \right )_{T} +  \left ( \frac{\partial \mathring{f}}{\partial t} \right )_{L}.
\end{cases}
\label{eq:transport2}
\eea
This should be sufficient for the purpose of understanding the leading as well as next-to-leading order phenomena at small energy densities. 

To describe the Nambu-Goto evolution of correlation vectors on long strings $\tilde{f}({\bf A}, {\bf B})$ we can still use (\ref{eq:nambu_long}), but the dynamics of correlation vectors on simple loops $\mathring{f}({\bf A}, {\bf B})$ should be slightly modified, i.e
\bea
 \left ( \frac{\partial \mathring{f}}{\partial t} \right )_{NG}  =   - \frac{\mathring{f}({\bf A}, {\bf B})}{A} - \frac{\mathring{f}({\bf A}, {\bf B})}{B} + \int d {\bf A'}  d {\bf B'}  \left ( \frac{1}{A A'\mathring{N}_A} +  \frac{1}{B B'\mathring{N}_B}  \right ) \mathring{f}({\bf A'}, {\bf B}) \mathring{f}({\bf A}, {\bf B'})  \delta_{({\bf A}+{\bf A'}-{\bf B}-{\bf B'})} \delta_{(A+A'-B-B')} \;\;\;\;\;\;
 \label{eq:nambu_loops}
\eea
where 
\bea
\mathring{N}_A({\bf A}, {\bf B})  \equiv \int  \frac{d {\bf A'} d {\bf B'}}{A'} \mathring{f}({\bf A'}, {\bf B'})  \delta_{({\bf A}+{\bf A'}-{\bf B}-{\bf B'})} \delta_{(A+A'-B-B')}  \notag\\
\mathring{N}_B({\bf A}, {\bf B})  \equiv \int  \frac{d {\bf A'} d {\bf B'}}{B'} \mathring{f}({\bf A'}, {\bf B'})  \delta_{({\bf A}+{\bf A'}-{\bf B}-{\bf B'})} \delta_{(A+A'-B-B')} .
\eea
Note that in contrast to the worldsheet of long strings described by (\ref{eq:nambu_long}), the world-sheet of loops is not generated with completely random diamonds but only with diamonds which can be on the same simple loops enforced by two delta functions $ \delta_{({\bf A}+{\bf A'}-{\bf B}-{\bf B'})}$ and  $\delta_{(A+A'-B-B')}$.

The transverse collisions between diamonds lead to sudden deformation of the worldsheet regardless of whether the participating diamonds belong to long strings or simple loops. Almost surely the aftermath of such collisions would be strings that are not describable with only four correlation vectors, resulting in a flow of energies from $\mathring{f}$ to $\tilde{f}$. Thus,  the transverse term derived above (\ref{eq:transverse}) can be easily generalized to simultaneously describe populations of simple loops
\bea
 \left ( \frac{\partial \mathring{f}}{\partial t} \right )_{T}  =  -p \int d {\bf A'}  d {\bf B'} \frac{\left ({A} + {B} \right )}{4} \left |\hat{A}\wedge\hat{B} \wedge \hat{A}'\wedge\hat{B}' \right |{f}({\bf A'}, {\bf B'}) \mathring{f}({\bf A}, {\bf B}) \;\;\;\;\;\;\;\;\;\;\;\;\;\;\;\;\;\;\;\;\;\;\;\;\;\;\;\;\;\;\;\;\;\;\;\;\;\;\;\;\;\;\;\;\;\;\;\;\;\;\;\;\;\;\;\;\;\;\;\;\;\;\;\;\;\;\;\;\;\;\;\;\;\;\; \label{eq:transverse_loops} 
 \eea
as well a long strings
\bea
 \label{eq:transverse_long}
 \left ( \frac{\partial \tilde{f}}{\partial t} \right )_{T}  =  -p \int d {\bf A'}  d {\bf B'}\frac{\left ({A} + {B} \right )}{4} \left |\hat{A}\wedge\hat{B} \wedge \hat{A}'\wedge\hat{B}' \right | {f}({\bf A'}, {\bf B'}) \tilde{f}({\bf A}, {\bf B})+ \;\;\;\;\;\;\;\;\;\;\;\;\;\;\;\;\;\;\;\;\;\;\;\;\;\;\;\;\;\;\;\;\;\;\;\;\;\;\;\;\;\;\;\;\;\;\;\;\;\;\;\;\;\;\;\;\;\;\;\;\;\;\;\;\;\;\;\;\;\;\;  \\
+ p \int d\alpha d {\bf A'}   d\beta d {\bf B'} \frac{\left ({A} + {B} \right )}{4} \left |\hat{A}\wedge\hat{B} \wedge \hat{A}'\wedge\hat{B}' \right |\left ( {f}({\bf A'}, \frac{\bf B}{\beta}) {f}(\frac{\bf A}{\alpha}, {\bf B'})  +  {f}(\frac{\bf A}{\alpha}, \frac{\bf B}{\beta}) {f}({\bf A'}, {\bf B'}) \right) \;\;\; \notag
 \eea
Note that the second line in (\ref{eq:transverse_long}) represents creation of infinitesimal segments and involves an integral of the energy density of all strings $f = \tilde{f} + \mathring{f}$. By substituting (\ref{eq:nambu_long}, \ref{eq:nambu_loops}, \ref{eq:transverse_long}, \ref{eq:transverse_loops}) in (\ref{eq:transport2}) we obtain a system of transport equations, i.e.
 \bea
 \label{eq:transport_system}
 \begin{cases}
\left ( \frac{\partial}{\partial t} +  {\bf v} \cdot \frac{\partial}{\partial {\bf x}} + \hat{\bf H} + \frac{1}{A} +  \frac{1}{B} \right ) \tilde{f} =    \int d {\bf A'}  d {\bf B'}   \int_{0 < \alpha,\beta < 1}  d\alpha d\beta  \left [ \left ( \frac{1}{A A' \tilde{N}_A} +  \frac{1}{B B' \tilde{N}_B}  \right ) \tilde{f}({\bf A'}, {\bf B}) \tilde{f}({\bf A}, {\bf B'})+ \right.\\ 
 \;\;\;\;\;\;\;\;\;\;\;\;\;\;\;\;\;\;\;\;\;\; \left. +   p \frac{\left ({A} + {B} \right )}{4} \left |\hat{A}\wedge\hat{B} \wedge \hat{A}'\wedge\hat{B}' \right | \left({f}({\bf A'}, \frac{\bf B}{\beta}) {f}(\frac{\bf A}{\alpha}, {\bf B'}) + \left ({f}(\frac{\bf A}{\alpha}, \frac{\bf B}{\beta}) -  \tilde{f}({\bf A}, {\bf B})  \right )  {f}({\bf A'}, {\bf B'}) \right )   \right ]+ \left ( \frac{\partial \tilde{f}}{\partial t} \right )_{L} \\
\left ( \frac{\partial}{\partial t} +  {\bf v} \cdot \frac{\partial}{\partial {\bf x}} + \hat{\bf H} + \frac{1}{A} +  \frac{1}{B} \right ) \mathring{f} =     \int d {\bf A'}  d {\bf B'} \left [   \left ( \frac{1}{A A'\mathring{N}_A} +  \frac{1}{B B'\mathring{N}_B}  \right ) \mathring{f}({\bf A'}, {\bf B}) \mathring{f}({\bf A}, {\bf B'}) \delta_{({\bf A}+{\bf A'}-{\bf B}-{\bf B'})} \delta_{(A+A'-B-B')}  \right . \\ 
 \;\;\;\;\;\;\;\;\;\;\;\;\;\; \;\;\;\;\;\;\;\;\;\;\;\;\;\;\;\;\;\;\;\;\;\;\;\;\;\;\;\; \;\;\;\;\;\;\;\;\;\;\;\;\;\;\;\;\;\;\;\;\;\;\;\;\;\;\;\;\;\;\;\;\;\;\;\;\;\; \left. - p \frac{\left ({A} + {B} \right )}{4} \left |\hat{A}\wedge\hat{B} \wedge \hat{A}'\wedge\hat{B}' \right | {f}({\bf A'}, {\bf B'}) \mathring{f}({\bf A}, {\bf B}) \right ]+ \left ( \frac{\partial \mathring{f}}{\partial t} \right )_{L},
\end{cases}
\eea
where only two longitudinal terms $\left ( \frac{\partial \tilde{f}}{\partial t} \right )_{L}$  and $\left ( \frac{\partial \mathring{f}}{\partial t} \right )_{L}$ remain to be calculated. 
   
In contrast to the transverse collisions which involve interactions of only two diamonds (or four correlation vectors), the longitudinal collisions must involve at least three nearby diamonds described with at least five correlation vectors. This is due to the fact that any pair of diamonds is allowed to undergo a longitudinal collision only if they are not nearest neighbors (e.g. E and NW  diamonds on Fig. \ref{fig:neighbors} can collide, but E and D cannot). In particular the collisions between next-to-nearest diamonds might produce a loop which is describable with only four correlation vectors in terms of the distribution function $\mathring{f}$. For example, the simplest longitudinal collisions takes place when a pair of left- moving correlation vectors collides with a triplet of right- moving correlation vectors or vise versa (see Appendix \ref{longitudinal} for details). Altogether the longitudinal production of loops is given by (\ref{eq:loops_longitudinal}), or
\bea
\label{eq:longitudinal_loops}
\left ( \frac{\partial \mathring{f}}{\partial t} \right )_{L} = p \int  \left ( \prod_{i=1,2,3} { d}\alpha_i {d}{\bf A}_i { d}\beta_i { d}{\bf B}_i \tilde{f}({\bf A}_i, {\bf B}_i)  \right ) \delta_{(\sum_{i=1,2,3} (\alpha_i {\bf A}_i - \beta_i {\bf B}_i ))} \delta_{(\sum_{i=1,2,3}( \alpha_i {A}_i - \beta_i {B}_i))} \times\;\;\;\;\;\;\;\;\;\;\;\;\;\;\;\;\;\;\;\;\;\;\;\;\;\;\;\;\;\;\;\;\;\;\;\;\; \\
\times \left [ \frac{\delta_{({\bf B}_1 - {\bf B}_2)} \delta_{(\alpha_2 -1)} \delta_{(\beta_2)}}{A_1 A_2 A_3 \tilde{N}_A^{(2)}}   \left (\delta_{({\bf A}-\alpha_1 {\bf A}_1-\alpha_3 {\bf A}_3)}\sum_{i=1,3} \delta_{({\bf B}-\beta_i {\bf B}_i)} \left ( \beta_i +\alpha_i \frac{A_i}{B_i}  \right )  +\delta_{({\bf A}-{\bf A}_2)} \delta_{({\bf B}-\beta_1 {\bf B}_2)} \frac{A_2}{B_2}  \right ) + ({\bf A} \leftrightarrow {\bf B})\right ] \notag
\eea
where $({\bf A} \leftrightarrow {\bf B})$ denotes the same as a previous term but with all of $\bf A$'s (i.e. ${\bf A}$, ${\bf A}_1$, ${\bf A}_2$ and ${\bf A}_3$) interchanged with all of $\bf B$'s (i.e. ${\bf B}$, ${\bf B}_1$, ${\bf B}_2$ and ${\bf B}_3$) . In addition to the longitudinal production of simple loops, the longitudinal collisions lead to a merger of correlation vectors on long strings resulting in growth of correlations described by (\ref{eq:long_longitudinal1},\ref{eq:long_longitudinal2}). Both integrals can be combine into the following longitudinal term
\bea
\label{eq:longitudinal_long}
\left ( \frac{\partial \tilde{f}}{\partial t} \right )_{L}  = p \int  \left ( \prod_{i=1,2,3} { d}\alpha_i { d}{\bf A}_i { d}\beta_i { d}{\bf B}_i \tilde{f}({\bf A}_i, {\bf B}_i)  \right ) \delta_{(\sum_{i=1,2,3} (\alpha_i {\bf A}_i - \beta_i {\bf B}_i ))} \delta_{(\sum_{i=1,2,3}( \alpha_i {A}_i - \beta_i {B}_i))} \times\;\;\;\;\;\;\;\;\;\;\;\;\;\;\;\;\;\;\;\;\;\;\;\;\;\;\;\;\;\;\;\;\;\;\;\;\; \\
\;\;\;\;\;\;\;\;\;\;\;\;\;\;\; \times\left [ \frac{\delta_{({\bf B}_1 - {\bf B}_2)} \delta_{(\alpha_2-1)} \delta_{(\beta_2)}}{A_1 A_2 A_3 \tilde{N}_A^{(2)}} \left ( \delta_{({\bf A}-(1-\alpha_1) {\bf A}_1-(1-\alpha_3) {\bf A}_3)} \sum_{i=1,3} \delta_{({\bf B}-(1-\beta_i) {\bf B}_i)} \left (1-\beta_i +(1-\alpha_i) \frac{A_i}{B_i}  \right )
 - \right. \right. \;\;\;\;\;\;\;\;\;\;\;\;\;\notag\\
\left. \left. -  \sum_{i=1,3}  \delta_{({\bf A}- {\bf A}_i)}  \delta_{({\bf B}- {\bf B}_i)} \left (1 +\frac{A_i}{B_i}  \right ) - \delta_{({\bf A}- {\bf A}_2)}  \delta_{({\bf B}- {\bf B}_2)} \frac{A_2}{B_2}   \right ) + ({\bf A} \leftrightarrow {\bf B})  \right] \notag.
\eea

This concludes our derivation of the transport phenomena for strings.  The system of transport equations (\ref{eq:transport_system}) together with longitudinal (\ref{eq:longitudinal_loops}, \ref{eq:longitudinal_long}) terms determine the leading as well as next to leading order dynamics of strings. Although the system might seem a bit complex, one can solve it fairly easily either numerically or perturbatively in various limits. In fact one such limit was already studied in \cite{Vanchurin2, Vanchurin3}, where the semi-scaling dynamics of cosmic strings was first analyzed. One can show that the transport equation for long strings (\ref{eq:transport_system}) reduces to an equation similar to Eq. (17) of Ref. \cite{Vanchurin3} in the limit where the longitudinal production of the smallest loops dominates the non-linear dynamics on long strings. 

\section{Conclusion}

The main objective of the paper was to derive transport equations which could describe a very complicated non-linear evolution of strings. We have shown that at large energy densities the transport phenomena can be describe with a single equation (\ref{eq:transport}), analogous to the Boltzmann transport equation for particles. The key ingredient in the construction was the distribution function $f({\bf A}, {\bf B}, {\bf x}, t)$ defined on a 9+1D phase space, which is sufficiently large to allow all of the relevant transport phenomena such as Nambu-Goto evolution, transverse collisions and gravitational effects. 

At small energy densities the analysis was further complicated by a necessity to simultaneously describe populations of long strings and small loops. This was accomplished by deriving a system of transport equations which is analogous to BBGKY hierarchy for particles. By truncating the hierarchy we have concentrated on a pair of coupled equations (\ref{eq:transport_system}) which is sufficient to describe leading as well as next-to-leading order transport phenomena (including longitudinal collisions) for long strings $\tilde{f}({\bf A}, {\bf B}, {\bf x}, t)$ as well as simple loops $\mathring{f}({\bf A}, {\bf B}, {\bf x}, t)$. 

The next step should be a careful analysis of the solutions of the transport equations, but it is already clear that the formalism developed in the paper should be useful in describing complicated non-linear systems with strings (e.g. fundamental strings, D-brane strings or field theory strings). For example the analysis of cosmic strings, which is usually tackled numerically due to non-trivial interaction between small and large scales,  should be greatly simplified with the help of derived equations. 

\section*{Acknowledgments}

The author is grateful to Shamik Banerjee, Xi Dong, Shamit Kachru, Shunji Matsuura, Mahdiyar Noorbala, Joe Polchinski, Steve Shenker, Eva Silverstein, Lenny Susskind and Alex Vilenkin for very helpful discussions. This work was supported in part by NSF Grant No.~0756174.

\appendix

\section{Friedmann background}\label{Friedmann}

The simple decomposition of the string solution into left- and right-moving waves (\ref{eq:wave}) is only valid in Minkowski space, but on other gravitational backgrounds the situation is more complex. As an example, consider the string motion ${\bf x}(\sigma,t)$ on a Friedmann background
\be
ds^2 = a(t)^2 (dt^2 - d{\bf x}^2) 
\ee
described by Hubble rate $H\equiv \frac{\dot{a}}{a}$ in conformal coordinates.  One can show that when two straight segments of opposite moving waves pass through each other the Hubble friction would produce interactions resulting in correlations of previously uncorrelated segments. Thus, our main task is to estimate the effect of such interactions on the correlation vectors $A$ and $B$ in comoving coordinates. 

From the equation of motion (\ref{eq:EOM}) the tangent vectors satisfy
\bea
\frac{d }{dt} {\bf a'} = H ({\bf b'} - ({\bf a'} \cdot {\bf b'}) {\bf a'})\\
\frac{d }{dt} {\bf b'} = H ({\bf a'} - ({\bf b'} \cdot {\bf a'}) {\bf b'})
\eea
and the comoving string length density evolves as
\be
\dot{{f}} =- H  2 {\bf v}^2  {f} .
\ee
(See Ref. \cite{BB} for details.) Then the correlation vectors defined in (\ref{eq:define_A}) must obey
\bea
\dot{\bf A} =  - H  \left \langle  \int_{-L/2}^{L/2} \left ({\bf b}'(\sigma) - ({\bf a}'(\sigma)  \cdot {\bf b}'(\sigma) ){\bf a}'(\sigma) \right ) \right \rangle = - C_1 H {\bf B}  - C_2 H (1- 2 {\bf v}^2) \hat{\bf A}
\label{eq:A_dot}
\eea
and
\bea
\dot{\bf B} =  - H  \left \langle  \int_{-L/2}^{L/2} \left ({\bf a}'(\sigma) - ({\bf b}'(\sigma)  \cdot {\bf a}'(\sigma) ){\bf b}'(\sigma) \right ) \right \rangle = - C_3 H {\bf A}  - C_4 H (1- 2 {\bf v}^2) \hat{\bf B}
\label{eq:B_dot}
\eea
where the factors $C_1$, $C_2$, $C_3$ and $C_4$ can be estimated as follows. The first term $-C_1H {\bf B}$ represents an action of the correlation vector ${\bf B}$ which is according to the integral of (\ref{eq:A_dot}) must takes place even if the segments $A$ and $B$  do not overlap. In other words ${\bf B}$ must be acting on the correlation vector ${\bf A}$ for it entire duration $\sim A$ even if segment $B$ acts on a given infinitesimal segment only during time interval $\sim B$. To take this effect into account we set $C_1 = A/B$ or similarly $C_3 = B/A$. In contrast to the first term, the second term $ - C_2 H (1- 2 {\bf v}^2) \hat{\bf A}$ does not vanish only when segment ${B}$ overlaps with segment ${A}$. The total world-sheet area of the overlap is $AB/2$ and the average time a given infinitesimal segment (which carries on the information about B) stays on $A$ is $A$, i.e. $C_2 = B/2$ or similarly $C_4=A/2$. Thus, 
\bea
\dot{\bf A} =  - H A \hat{{\bf B}} - H B \hat{\bf A} \frac{1- 2 {\bf v}^2}{2}
\eea
and
\be
\dot{\bf B}  = - H B \hat{ {\bf A}}  - H A \hat{{\bf B}} \frac{1- 2 {\bf v}^2}{2}.
\ee
The resulting operator
\be
\hat{\bf H} \equiv H \left ( 2 {\bf v}^2  -  B \hat{\bf A} \cdot \left ( \frac{1- 2 {\bf v}^2}{2} \frac{\partial}{\partial {\bf A}} +  \frac{\partial}{\partial {\bf B}} \right ) -  A \hat{\bf B} \cdot \left ( \frac{1- 2 {\bf v}^2}{2} \frac{\partial}{\partial {\bf B}} +  \frac{\partial}{\partial {\bf A}} \right ) \right ).
\label{eq:grav}
\ee
describes the evolution of correlation vectors in the transport equation (\ref{eq:transport}) under assumption of negligible back-reaction. Although we have only calculated the gravitational effects for Friedmann backgrounds it should be relatively easy to generalize the formalism to more complicated backgrounds, but is not immediately clear how to include back-reaction which lies beyond our semi-classical analysis. 

\section{Longitudinal collisions}\label{longitudinal}

\begin{figure}[htbp]
   \begin{center}
   \includegraphics[width=5in]{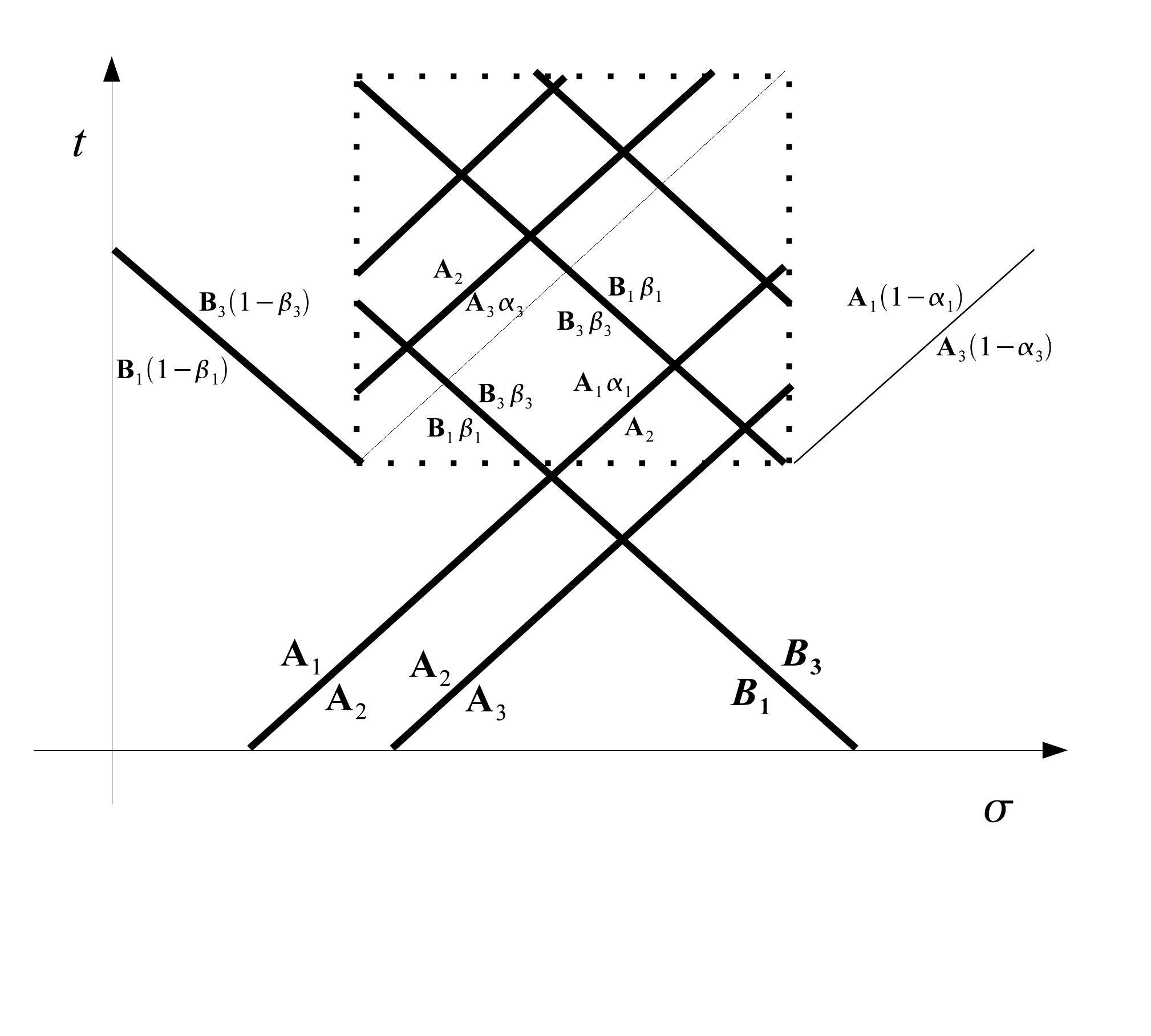}
   \caption{Longitudinal collision of two left-moving correlation vectors with three right-moving correlation vectors.}
   \label{fig:longitudinal}
   \end{center}
   \end{figure}

\begin{figure}[htbp]
   \begin{center}
   \includegraphics[width=4in]{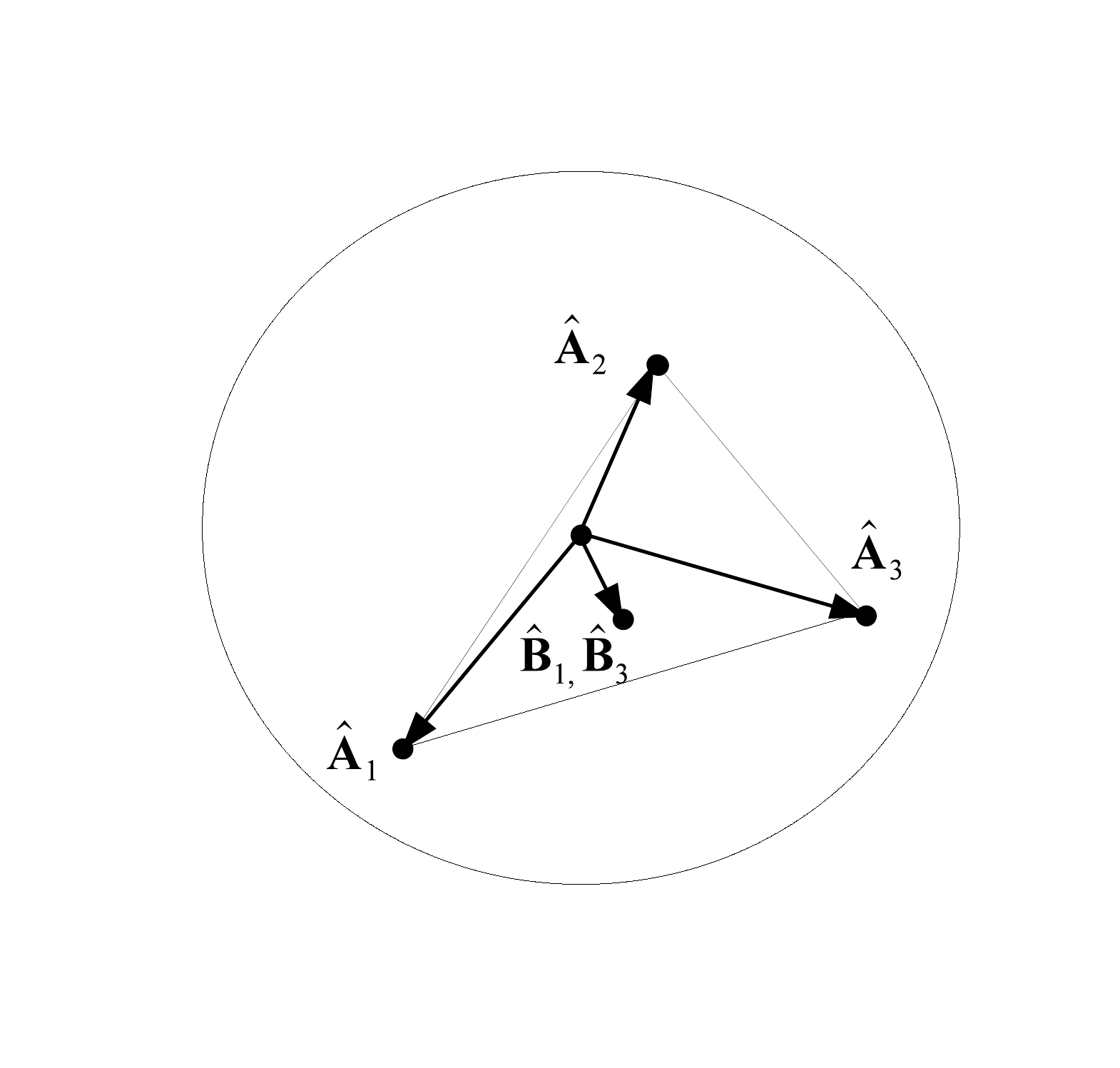}
   \caption{Projection of unit vectors $\hat{\bf A}_1, \hat{\bf A}_2, \hat{\bf A}_3, \hat{\bf B}_1, \hat{\bf B}_3$  on a plane orthogonal  to $\hat{\bf B}_1- \hat{\bf B}_3$.}
   \label{fig:unit_circle}
   \end{center}
   \end{figure}

 Consider a longitudinal collision between two left-moving correlation vectors $\bf B_1$ and $\bf B_3$ and three right-moving correlation vectors $\bf A_1$, $\bf A_2$ and $\bf A_3$ (see Fig. \ref{fig:longitudinal}).  Under the string chaos assumption the directions of these correlation vectors are completely random but not all of such regions would emit simple loops. To distinguish the configurations that could lead to production of loops from those that cannot and to understand the aftermath of such collisions it is useful to plot a projection of unit vectors  $\hat{\bf A}_1, \hat{\bf A}_2, \hat{\bf A}_3, \hat{\bf B}_1, \hat{\bf B}_3$ on a plane orthogonal to $\hat{\bf B}_3- \hat{\bf B}_1$ (see Fig. \ref{fig:unit_circle}).
It is easy to check that for sufficiently large $A_1, A_3, B_1, B_3$ a loop of size $\gtrsim A_2$ would form whenever the $\hat{\bf B}_1, \hat{\bf B}_3$ point lies inside of the triangle with vertices in  $\hat{\bf A}_1, \hat{\bf A}_2$ and $\hat{\bf A}_3$. And if so, then the vectors $\hat{\bf A}_1$ and $\hat{\bf A}_3$ must be correlated with each other. (Such correlations are represented on Fig.  \ref{fig:longitudinal} with a finer line along the kink connecting ${\bf A}_1 \alpha_1$ and ${\bf A}_3\alpha_3$.) Thus, in order for the string chaos assumption to remain valid we will approximate such loops with only four correlation vectors ${\bf B}_1 \beta_1$, ${\bf B}_3\beta_3$, ${\bf A}_2$ and ${\bf A}_1\alpha_1 + {\bf A}_3\alpha_3$ which could be described in terms of the distribution function $\mathring{f}$. 

\begin{figure}[htbp]
   \begin{center}
   \includegraphics[width=4.5in]{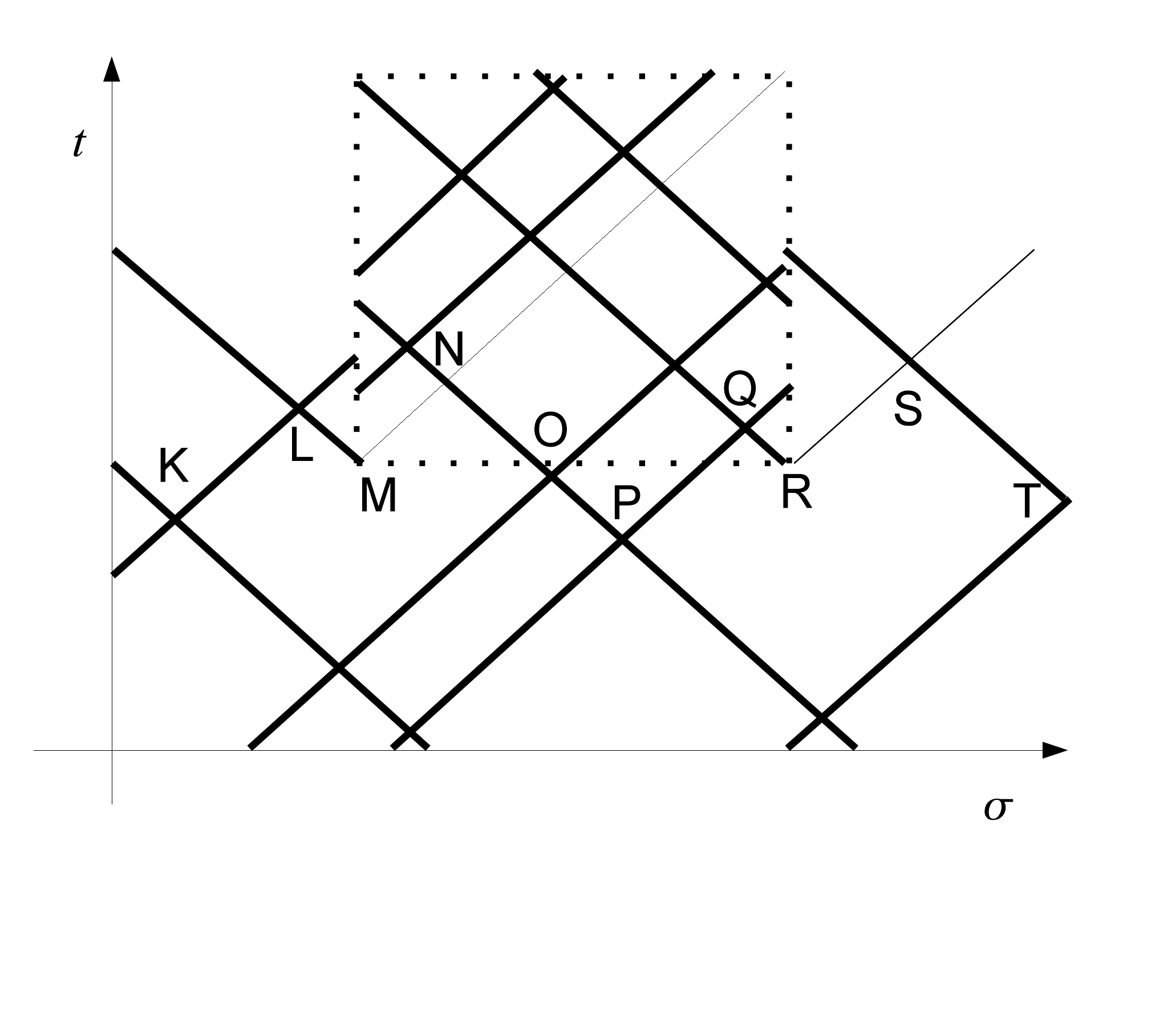}
   \caption{The boundaries $KL, LM, MN, NO, OP, PQ, QR, RS$ and $ST$ describe possible transitions of infinitesimal segments during longitudinal collisions.}
   \label{fig:longitudinal2}
   \end{center}
   \end{figure}

It is convenient to analyze the transitions of infinitesimal segments though boundaries $KL, LM, MN, NO, OP, PQ, QR, RS$ and $ST$ (see Fig.\ \ref{fig:longitudinal2})  
with separate integral expressions. In particular the transitions across $MN, NO, OP, PQ, QR$ would describe a flow of energy from $\tilde{f}$ to $\mathring{f}$ and transitions across $KL, LM, RS, ST$ describe a flow of energy from $\tilde{f}$ to $\tilde{f}$. The corresponding creation of infinitesimal segments along boundaries $MN, NO, OP, PQ, QR$ is given by the following integral:
\bea
\label{eq:loops_longitudinal}
\int  \left ( \prod_{i=1,2,3} { d}\alpha_i { d}{\bf A}_i { d}\beta_i { d}{\bf B}_i \right ) \tilde{f}({\bf A}_1, {\bf B}_1)   \tilde{f}({\bf A}_2, {\bf B}_2)  \tilde{f}({\bf A}_3, {\bf B}_3) \; \delta_{(\sum_{i=1,2,3} (\alpha_i {\bf A}_i - \beta_i {\bf B}_i ))} \delta_{(\sum_{i=1,2,3}( \alpha_i {A}_i - \beta_i {B}_i))} \times\;\;\;\;\;\;\;\;\;\;\;\;\;\;\;\;\;\;\;\; \\
\frac{\delta_{({\bf B}_1 - {\bf B}_2)} \delta_{(\alpha_2 -1)} \delta_{(\beta_2)}}{A_1 A_2 A_3 \tilde{N}_A^{(2)}}   \left (\delta_{({\bf A}-\alpha_1 {\bf A}_1-\alpha_3 {\bf A}_3)}\sum_{i=1,3} \delta_{({\bf B}-\beta_i {\bf B}_i)} \left ( {\beta_i} +\alpha_i \frac{A_i}{B_i}  \right )  +\delta_{({\bf A}-{\bf A}_2)} \delta_{({\bf B}-\beta_1 {\bf B}_1)} \frac{A_2}{B_1}  \right ) \notag
\eea
where 
\be
 \tilde{N}^{(2)}_A = \int { d} {\bf A}_1 { d}{\bf B}_1 { d}{\bf A}_2   \frac{\tilde{f}({\bf A}_1, {\bf B}_1)\tilde{f}({\bf A}_2, {\bf B}_1)}{A_1 A_2}; \;\;\; \;\;\; \tilde{N}^{(2)}_B = \int { d} {\bf A}_1 { d}{\bf B}_1 { d}{\bf B}_2   \frac{\tilde{f}({\bf A}_1, {\bf B}_1)\tilde{f}({\bf A}_1, {\bf B}_2)}{B_1 B_2}.
 \ee
For example the term with $\delta_{({\bf A}-\alpha_1 {\bf A}_1-\alpha_3 {\bf A}_3)}$ describes transitions across $MN, NO, PQ, QR$ boundaries and the term with $\delta_{({\bf A}-{\bf A}_2)}$ describes the transitions across $OP$ boundary.  Similarly, the corresponding creation of infinitesimal segments on boundaries  $KL, LM, RS, ST$ is given by
\bea
\label{eq:long_longitudinal1}
\int  \left ( \prod_{i=1,2,3} { d}\alpha_i { d}{\bf A}_i { d}\beta_i { d}{\bf B}_i \right ) \tilde{f}({\bf A}_1, {\bf B}_1)   \tilde{f}({\bf A}_2, {\bf B}_2)  \tilde{f}({\bf A}_3, {\bf B}_3) \; \delta_{(\sum_{i=1,2,3} (\alpha_i {\bf A}_i - \beta_i {\bf B}_i ))} \delta_{(\sum_{i=1,2,3}( \alpha_i {A}_i - \beta_i {B}_i))} \times\;\;\;\;\;\;\;\;\;\;\;\;\;\;\;\;\;\;\;\; \\
\times \left ( \frac{\delta_{({\bf B}_1 - {\bf B}_2)} \delta_{(\alpha_2 -1)} \delta_{(\beta_2)}}{A_1 A_2 A_3 \tilde{N}_A^{(2)}}  \delta_{({\bf A}-(1-\alpha_1) {\bf A}_1-(1-\alpha_3) {\bf A}_3)} \sum_{i=1,3} \delta_{({\bf B}-(1-\beta_i) {\bf B}_i)} \left ( 1-\beta_i +(1-\alpha_i) \frac{A_i}{B_i}  \right ) \right ) \notag
\eea
and the overall destruction of infinitesimal segments on all boundaries ($KL, LM, MN, NO, OP, PQ, QR, RS, ST$) is given by 
\bea
\label{eq:long_longitudinal2}
-\int  \left ( \prod_{i=1,2,3} { d}\alpha_i { d}{\bf A}_i { d}\beta_i { d}{\bf B}_i \right ) \tilde{f}({\bf A}_1, {\bf B}_1)   \tilde{f}({\bf A}_2, {\bf B}_2)  \tilde{f}({\bf A}_3, {\bf B}_3) \; \delta_{(\sum_{i=1,2,3} (\alpha_i {\bf A}_i - \beta_i {\bf B}_i ))} \delta_{(\sum_{i=1,2,3}( \alpha_i {A}_i - \beta_i {B}_i))} \times\;\;\;\;\;\;\;\;\;\;\;\;\;\;\;\;\;\;\;\; \\
\times \frac{\delta_{({\bf B}_1 - {\bf B}_2)} \delta_{(\alpha_2 -1)} \delta_{(\beta_2)}}{A_1 A_2 A_3 \tilde{N}_A^{(2)}}   \left (\sum_{i=1,3} \delta_{({\bf A}- {\bf A}_i)}  \delta_{({\bf B}- {\bf B}_i)} \left (1+\frac{A_i}{B_i}  \right ) + \delta_{({\bf A}- {\bf A}_2)}  \delta_{({\bf B}- {\bf B}_2)} \frac{A_2}{B_2}  \right ).  \notag
\eea

The three integrals (\ref{eq:loops_longitudinal}, \ref{eq:long_longitudinal1}, \ref{eq:long_longitudinal2}) describe longitudinal collisions of two left-moving correlation vectors with three right-moving correlation vectors, but similar expressions would also hold for collisions of two right-moving vectors with three left-moving vectors with all of $\bf A$'s interchanged with all of $\bf B$'s.  If desired one can also include more complicated longitudinal collisions involving more than five correlation vectors which would correspond to higher order effects.

\end{document}